\documentclass[preprint2]{aastex}
\usepackage{amsmath} 

\newcommand{\parenthnewln}{\right.\\&\left.\quad\quad{}}

\shorttitle{Observing Planet Destruction}
\shortauthors{Taylor}

\begin{document}


\title{Observing the Luminosity Increase and Roche Lobe Overflow of Planet Destruction}


\author{Stuart F. Taylor\altaffilmark{1,2} }
\affil{Department of Physics and Institute of Astronomy, 
    National Tsing Hua University, Hsinchu, Taiwan}


\altaffiltext{1}{Global Telescope Science, Los Angeles, California, USA}
\altaffiltext{2}{Eureka Scientific, Oakland, California, USA}


\begin{abstract}
The destruction of planets by migration into the star will release significant amounts of energy and material, which will present opportunities to observational study planets in new ways. To observe planet destruction, it is important to understand the processes of how this energy and material is released as planets are destroyed. It is not known how fast the large amounts of energy and material are released, making it difficult to predict how observable planet destruction will be. There is a huge amount of energy made available by falling deep into the star's potential well: Simple calculations show that many of the currently known``hot Jupiters'' can potentially produce events as luminous as a small nova if the energy is released fast enough. To observe these events, the important questions are how will this energy be released, and whether the energy will be released rapidly enough to create an event luminous enough to be found by transient surveys. Even the final rapid tidal infall of the planet may input enough energy into the star that could rival the energy output of the star. The final destruction will release even far more energy. These events are rare enough to explain why there is no undisputed observation of planet accretion, but future transient astronomy surveys such as PTF, Pan-STARRS, and LSST may have a good chance of catching planet destruction events. 

Alternatively, if planet destruction is slowed by the inward migration alternating with periods of outward migration caused by Roche lobe overflow (RLOF), then the primary signature may be the effects of the release of large amounts of gas. The infall of this gas also may significantly contribute to the system's luminosity. The release of planetary gas may be a searchable signature of planet destruction. Signs of runaway RLOF and outward or alternating RLOF should be searched for. RLOF unevenly distributed through the orbit may maintain an eccentricity through orbitally asymmetric transfer of angular momentum.

Smaller planets may provide less dramatic but potentially more common events. Planet formation models frequently refer to protoplanet infall. Comparison with solar system objects shows how these events may be much brighter than seen in ordinary YSO and proto-star variability.

\end{abstract}




\section{Introduction}

Planets can be destroyed not only during planetary formation and when stars expand, but also during the entire lifetime of stars including during the main sequence. How the energy and gas release of these ``end states'' of planets it still not well understood, though much important work has been started (Jackson et al. 2009; Levrard et al. 2009; Li et al. 2010; Fossati et al. 2010; Gu et al. 2003; Chang et al. 2010; Jiang et al. 2003; Trilling et al. 1998; Shaviv \& Starrfield 1988; Siess \& Livio 1999a,b; Soker \& Tylenda 2003.) Much recent attention has been on exoplanets migrating towards their stars. Theoretical work on planet destruction (eg. Jackson et al. 2009;  Levrard et al. 2009; Li et al. 2010) has been supported by observations of gas release through Roche lobe overflow (RLOF) that could be considered as the ``beginning of the end'' of the most close-in planets (Fossati et al. 2010). The closest exoplanets now being found close to their stars may be tidally spiraling in to their destruction even during their main sequence. Though planet consumption has long been expected to happen late in the evolution due to explanion of the star, and has long been considered during planet formation, the earlier consensus was that such planets were not likely to migrate into the star during the main sequence phase. Tidal migration has become one of the fastest developing areas of exoplanet study, with extraordinarily rapid exchange of new understandings from new observational and theoretical results. Within the last few years there have been enough transiting exoplanets for patterns of tidal migration to become apparent through study of orbital parameters. New theoretical results and observations have brought new insights addressing the question of how the closest in planets get destroyed by tidally migrating into their stars. Theoretically, that the destruction of planets by inspiral into the star actually must occur has been shown by recent work that considers the effect of tides induced on the star (eg: Jackson et al. 2009;  Levrard et al. 2009; Chang et al. 2010), following on earlier work that addressed planet infall even during the star's main sequence phase (Gu et al. 2003; Rasio et al. 1996a,b). Early work on the formation and migration of giant planets showed that migration in the disk could send planets into the star (Masset \& Papaloizou 2003; Lin \& Papaloizou 1986).  The detection of a few unexpectedly close massive planets such as WASP-18b (Hellier et al. 2009) has challenged  tidal migration theory to explain the statistics of finding planets with such small semi-major axes, but it appears that planets must continue to migrate towards the star long after formation (Hellier et al. 2009), and thus are likely to be accreted even after the dissipation of the circumstellar disk. While many other questions regarding tidal theory are being addressed, it has not been fully considered how much energy these massive planets will deliver to their stars. Planet consumption has been considered for causing  contamination of the star by planetary material (Santos et al. 2010; Sandquist et al. 1998; Sandquist et al. 2002), even including planet contamination of white dwarfs (Jura et al. 2009), and the effect of stellar spinup have been considered (Pont 2009; Livio \& Soker 2002). Planet consumption  during late-stage stellar expansion has been widely considered (Soker \& Tylenda 2003), including as an explanation for the outburst of V838 Monocerotis (hereafter V838 Mon; Retter \& Marom 2003) and for features of planetary nebulae (Soker 1996). These works included consideration of the large energy of such consumption (Siess \& Livio 1999), and work on the energy release has been done for the case of accretion onto a white dwarf (Shaviv \& Starrfield 1988). While this work has been done on energy release from the planet occurs, it is not clear how planet accretion energy will be radiated in most of the many cases. 

Much work has been done on the possibility that planets  have contaminated stars (Israelian 2001, 2004; Ashwell et al. 2005) and even white dwarfs (Farihi et al. 2010a;  Farihi et al. 2010b; Jura et al. 2009a; Jura et al. 2009b) with Li isotopes and heavier elements. It is important to follow through the stages of the end states of planets. 

Planet destruction has long been considered to occur when a star expands (Shaviv \& Starrfield 1988). It has also been proposed that the brightenings of V838 Mon and a small number of similar stars in nearby galaxies might be explained by the consumption of planets (Soker \& Tylenda 2003). The controversy over these events may illustrate the difficulty in confirming a particular brightening to be a planet destruction event. V838 had three brightenings requiring either the consumption of three planets, or three separate times of mass loss for a single planet. If V383 Mon was a planet consumption event, we would expect a significant fraction of similar brightenings to be explainable by the consumption of single planets of different masses. The limited time-series coverage of other candidate events M31-RV and V4332 Sgr (Bond \& Siegel 2006) illustrate the importance of having improved data from upcoming time surveys with followup programs. We are not focused on planet consumption due to the star's evolution due to the previous work on planet consumption by evolved stars. However, planet consumption that is due as much to the star expanding as to tidal migration may be much more likely to avoid RLOF and produce a bright event, when the outer layers of the star expand to distances at which the planet does not RLOF. 

In Section 2 we address the patterns that lead us to consider planet destruction as an important event (section 2.1), and given these patterns how often it may occur (section 2.2). In Section 3 we emphasize the large amount of energy that can be released, and address the major question of how much of this energy will be released during RLOF or collision. In Section 4 we discuss how the evolution of tidal migration. In Section 4.1 we describe the star's rotation as the controlling factor allowing stellar tides to  cause tidal inspiral, but in Section 4.2 we consider that magnetic braking can slow stellar rotation. We describe how RLOF can change the inward migration to alternating inward/outward migration. In Section 5 we briefly discuss the inspiral of smaller bodies during the planet formation phase. In Section 6 we advocate future work that can be done during already planned transient astronomy observations that may find these events. In Section 7 we conclude with an emphasis on observing not only the energy release but also seeking to observe the material coming off planets that will be a sign of planet-destruction related RLOF.

\section{Patterns in planet destruction}

In section 2.1 we address how the distribution of planets may be better explained by having lost planets to inspiral into the star, and in section 2.2 we comment on how a luminous signature might be rare enough to not have been previously identified, but not so rare that current and upcoming transient surveys may catch potential events.

\subsection{Planets gone missing}

A major puzzle has been the apparent ``pile-up'' of planets at periods of two or three days, with the plausible explanations for bringing planets inward including Kozai oscillations, secular migration, and planet-planet scattering (Mazeh 2008; Wu et al. 2007; Faber et al. 2005). So called ``Hot Jupiters'' have long been thought to have migrated to their current positions from a more distant formation location. Previous efforts to solve this puzzle sought explanations that have the planet absolutely stopping short of falling into the star, and indeed, some of the processes such as Kozai oscillations do not continue to operate all the way to bringing a planet into a star (Fabrycky \& Tremaine 2007; Wu et al. 2007), though there is no reason other than the small cross section that planet-planet scattering could not move a planet into a collision orbit with the star. Jackson et al. (2009) and Levrard et al. (2009) showed that once these other process place planets in this pileup, that tides caused by the planet on the star may be large enough  to continue the inward migration until the planet is destroyed either by RLOF or collision. The rate of tidal migration depends on the poorly known amount of stellar tidal energy dissipation, which is represented in inverse form by the ``tidal quality factor'' or 
``$Q$'' 
value. The question of how much energy is dissipated, leading to what value of $Q_{\ast}$, has been the subject of much research which is leading to a consensus that $Q_{\ast}$ is likely a more complicated function than previously assumed. Though these tides are still small in the pileup, the tides on the star are such a stiff function of semi-major axis that they may  help define the inner edge of the pileup (Jackson et al. 2009; Levard et al. 2009; Ibgui \& Burrows 2009).  Despite the plausible reasons for migration to slow on approach to the star in a manner that may explain this pile-up (Wu et al. 2007), these reasons do not require that the migration stop. We emphasize downward migration of the semi-major axis, but note that some processes that affect the eccentricity such as planet-planet interactions do not necessarily have a barrier preventing eccentricity to evolve so close to 1 that the planet could collide with the star, releasing slightly more energy than inspiralling planets. (Fabrycky \& Tremaine [2007] show that Kozai oscillations do have a barrier preventing collision with the star). We emphasize inspiral into the star, which we assume to be much more common.

It had been assumed that after the circumstellar disk dissipates, planets were considered to primarily migrate due to tides raised by the star onto the planets. Since the star can raise variable tides on planets only if when the planet´s orbit is eccentric, it was shown that since the tides circularize the planets orbit that the planets will indeed be preferentially deposited in orbits of two or three days. The assumption that planets forever remain in these orbits was challenged by Jackson et al. (2008a,b,c,d) and Levrard et al. (2009), who noticed that at these close-in orbits, the tides raised by the planets had been wrongly neglected. Not only do planets in three day orbits raise enough tides on the star to cause continued slow migration in even day-long close-in orbits, but this tidal migration is such a stiff function of semi-major axis that the inner edge of this pileup may be determined by planets migrating out of this pileup into the star. Jackson et al. (2009) pointed out that it may be that the inner age of this pileup may be a function of stellar age, as more of the ``piled up'' planets further out migrate into the star. Selection effects and tidal spinup may confound the result as originally presented (Jackson, personal communication), so improved evidence of a stellar age-dependence of minimum planet semi-major axis is worth seeking out. Jackson et al. (2009) also list several likely observational consequences of planets likely migrating into the star, including searching for rapidly rotating stars and stars contaminated by planetary metallicities.

\subsection{Planets gone missing}

If planet destruction of highly massive planets is a highly luminous event, it would be rare enough to explain why there are no claims of observing planet destruction due to inspiral. The only possible claims of observing planet destruction events are for when the star expands and consumes the planet, as in the stars brightening similar to V838 Mon (e.g. Retter \& Marom, 2003). Combining the frequencies of ``Very Hot Jupiters'' and ``Hot Jupiters'' given by Gould et al. (2006) gives one in 210 stars that currently hosts a planet that may create such an event, but it is unknown how many stars have already lost planets. If the current star formation rate of the galaxy is 7 stars per year (Diehl et al. 2006), then such an event may occur once every 30 years per galaxy, and possibly more frequently depending on what fraction of the original number of planets the current population represents. This is a low enough rate to require the monitoring of several nearby galaxies by transient sky surveys, something that is being started with PTF and Pan-STARRS, and will be done with greater coverage with LSST (Beatty \& Gaudi et al. 2008).

The key question for observing luminous events at these rates is whether the events will be fast and luminous, or if the gas will be bled off slowly enough to not produce dramatic events. The possibilities that will lead to luminous events would be if enough of a dense part of the planet survives and impacts with the star either during main sequence inspiral or when the star expands, and a sufficiently rapid runaway RLOF. A slower RLOF could be much less luminous, but could allow for more opportunities to find close-in planets releasing sufficiently large amounts of gas that will make these objects prime observing targets.

\section{Significant energy release}
\begin{figure}
\epsscale{.80}
\plotone{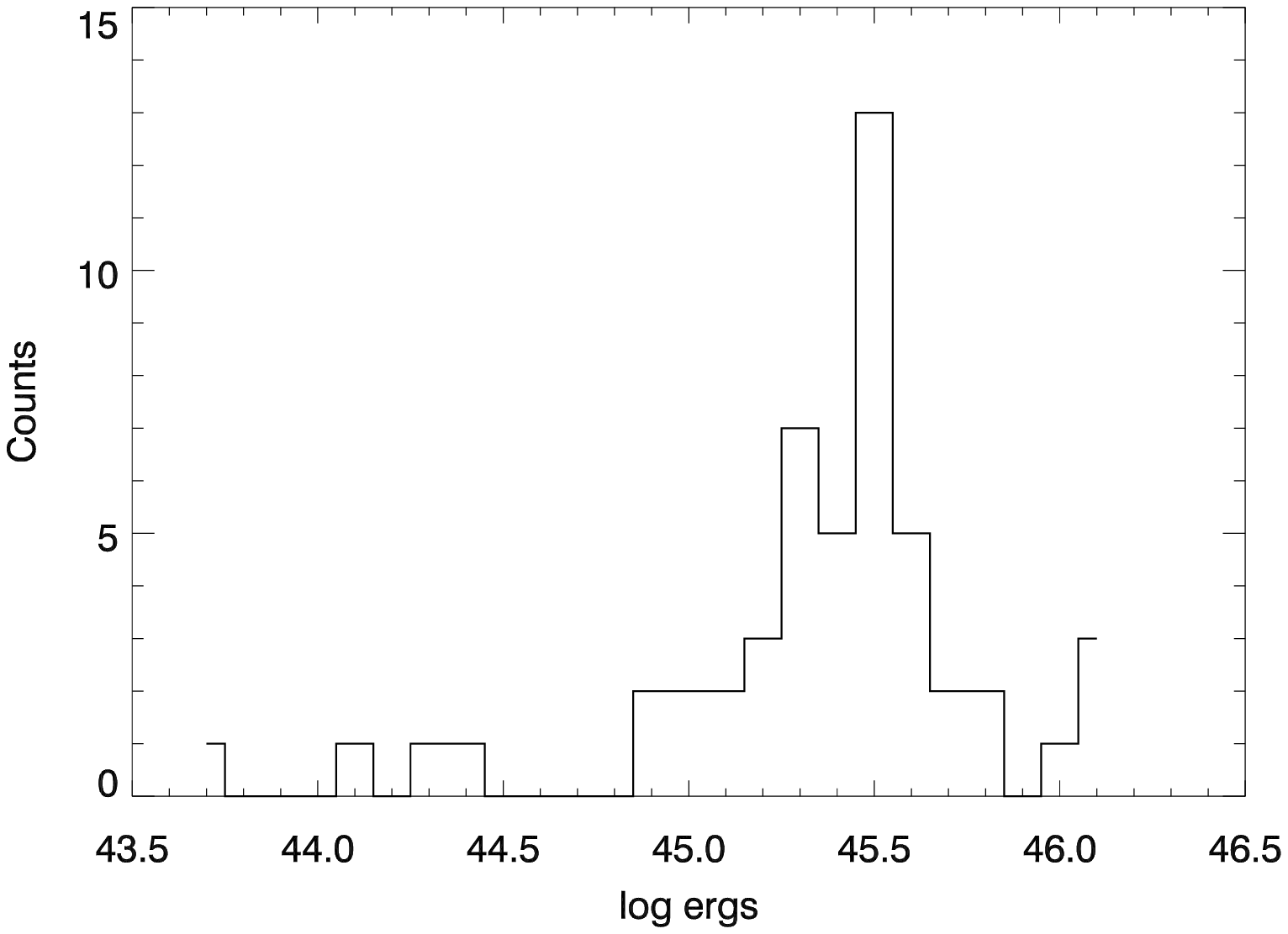} 
\caption{Log of orbital energy (kinetic plus potential) from current orbital semimajor axis of the transiting planets closer than 0.05 AU with complete parameters (listed in the Exoplanet Encyclopedia [exoplanet.eu] as of 2010 August).}\label{ergstart}
\end{figure}

\begin{figure}
\epsscale{.80}
\plotone{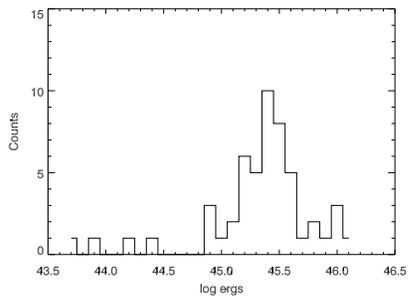} 
\caption{Log of orbit energy from point of planet destruction. This is defined to be when RLOF stars or when planet collides with stellar photosphere. Comparison with Fig~ref{ergsstart} shows that most of the energy is released after the planet has started to be destroyed.}\label{ergend}
\end{figure}

\begin{figure}
\epsscale{.80}
\plotone{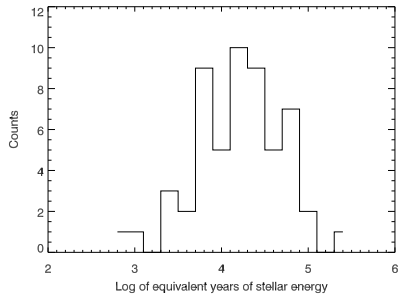} 
\caption{Log of ``equivalent years of stellar energy'' from current orbit. This is the number of years of each star's luminosity that is equivalent to the orbital energy.}\label{yearsstart}
\end{figure}

\begin{figure}
\epsscale{.80}
\plotone{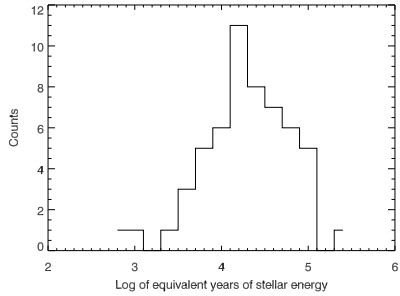} 
\caption{Log of equivalent years of stellar energy from point of destruction as defined in caption to Figure~\ref{ergend}. Comparison with Figure~\ref{yearsstart} shows that again, after the energies have been rescaled according to stellar luminosities, that most of the energy is released after the planet has started to be destroyed.}\label{yearsend}
\end{figure}

\begin{figure}
\epsscale{.80}
\plotone{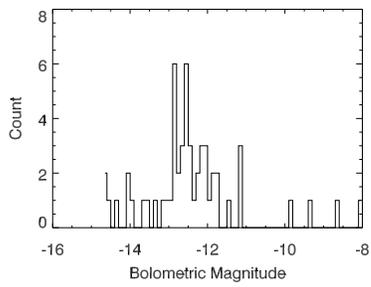} 
\caption{The bolometric luminosity for an example one 24-hour day release of the orbital energy.}\label{bolometric}
\end{figure}

The orbital energies of the most massive planets is equal to tens of thousands of years of stellar luminosity. We plot the orbital energies (potential plus kinetic) of the transiting planets closer than 0.05 AU  with full orbital parameters and temperatures available 2010 August 1 in the Extrasolar Planet Encyclopedia (exoplanet.eu). 
We calculate the current and ``final'' orbital energies by adding the orbital kinetic energy to the potential energy relative to the stellar photosphere (one stellar radius). We define the final energy as where what we call the start of  ``planet destruction'' to occur. We consider ``planet destruction'' to be occuring either at the start of RLOF or if the planet undergoes collision with the photosphere. 
The energies are of the order of $10^{45}$ erg as illustrated in Figure~\ref{ergstart}. This energy is of the order of a small nova. We see in Figure~\ref{ergend} that the final orbital energies are not much less than the current energies, meaning most of this energy is released after the start of planet destruction. 
By dividing these energies by the yearly luminosity energies of each host star,
we obtain Figure~\ref{yearsstart}, which shows that the energy available is the equivalent of the order of $10^4$ to $10^5$ years of stellar output. Expressed this way, we again see in Figure~\ref{yearsend} that most of this energy is release during the process(es) of planet destruction. We also can see that if this energy is released quickly, then these events have the potential of being detected at great distances. In Figure~\ref{bolometric} we see what the bolometric magnitude would be if the energy at RLOF or collision distances were released in one day. A real event could release its orbital energy moderately more slowly and thus be much less bright than this, but still be observable in nearby galaxies by upcoming surveys, especially by LSST, which will observe to better than magnitude 18.5 (Beatty \& Gaudi 2008). Less massive planets may slowly release their energy through RLOF. The wide range of planet accretion events will vary, both with planet mass and with when in the stages of planet system evolution the destruction occurs.

These large energies are enough to to create very luminous events if the energy is released quickly. To demonstrate this, we show in Figure~\ref{bolometric}
the brightness that these events could be if the energy were all released evenly through one 24-hour day. It is neither expected that the energy will be released this fast nor that the energy release would necessarily be constant, but this arbitrary choice is convenient to think of observationally and to scale down from. Even if these optimistic brightnesses are scaled down some, these brightnesses raise the possibility that current transient surveys such as the Palomar Transient Factory (PTF) and Pan-STARRs may catch such events in M-31, and the LSST could monitor a few more of the nearest Milky-Way size galaxies for such events. The main challenge may be the rarity of such events (discussed below).
We advocate work addressing the questions of how transient surveys might identify observations of planet destruction in both the tidal migration and final destruction stages. During tidal migration, how is the significant energy release of massive planets migrating down many stellar radii released? How does disk extinction affect observations of protoplanet accretion? Final destruction will occur through different density-dependent channels, from RLOF of small planets to orbital merging with the stellar photosphere for the most massive planets. The process of inward migration will be slowed by the planet synchronizing the rotation of the star leaving less difference between the rotations of the orbit and the star. While the speedup of the star's rotation will be reduced or reversed by magnetic braking (Barker \& Ogilvie 2009; Dobbs-Dixon et al. 2004). Matsumura et al. (2010) find in their simulation of the orbits of transiting planets that even though magnetic braking may, in some cases, slow the evolution of these orbits, the changes are small enough to not greatly affect results such as how planet destruction proceeds or how often planet destruction might occur. If the final destruction energy is released quickly enough, upcoming surveys may catch the brightest of these events even in nearby galaxies. 
The challenge will be to identify when and how this higher luminosity will produce an observable signature.  It will be important to identify signatures identifiable to upcoming transient surveys.  We advocate efforts to model and observe what we expect will include the most energetically dramatic events possible in solar systems evolution.

During the final infall, for the most massive planets the energy from tides may also become significant relative to the energy output of the star, depending on the rate of tidal energy dissipation in the star, which is directly related to the stellar tidal quality factor $Q_{\ast}$. 
We calculate the energy dissipation to be the energy lost to tides on the star,
which is the negative of the rotational energy change of the system
(Murray \& Dermott, 1999). The
change of energy of the system is the change in the orbital energy minus the
energy input into speeding up the rotation of the star. We neglect the small
amount of energy input into speeding up the rotation of the planet.

We relate this energy change to the change in semi-major axis. The equations
of the change in semi-major axis that we use 
include the change of rotation, and are also important in Section 4.1
where we address how the tidal migration rate is coupled to stellar spin-up.
To do this we follow Levard et al. (2009) in using Neron de Surgy \& Laskar's 
(1997) expression that gives the change in semimajor axis to
second order in eccentricity for the evolution of the semi-major axis,
assuming zero planetary obliquity.
\begin{align}\label{levardeq3} 
\nonumber \frac{da}{dt}= 
& \frac{6 M_p R^5_{\ast}}{Q^{\prime}_{\ast} M_{\ast} a^4}
\left
  [(1+\frac{27}{2}e^2){\omega}_{\ast} \cos \varepsilon 
  -(1+23 e^2) n 
\right] \\ 
 & + 
\frac{6 M_{\ast} R^5_p}{Q^{\prime}_{p} M_{p} a^4}
\left
  [(1+\frac{27}{2}e^2) {\omega}_{p}
  -(1+23 e^2) n 
\right] ,
\end{align} 
where $\omega_{\ast}$ and  $\omega_p$ are the 
stellar and planetary rotation velocities respectively. $\varepsilon$ is the
stellar obliquity, $a$ is the semimajor axis, $e$ is the eccenctricity, 
and $n$ is the orbital mean motion. 
$Q^{\prime}_{\ast}$ and $Q^{\prime}_{p}$ are the ratios
between the present annual stellar and planetary tidal quality factors 
$Q_{\ast}$ and $Q_{p}$, and the tidal Love number of degree 2 
$k_{2,\ast}$ and $k_{2,p}$ respectively. The first term, with 
$Q^{\prime}_{\ast}$ in the denominator, is due to tides within the star, 
and the second term, with $Q^{\prime}_{p}$ in the denominator, is due
to tides within the planet.
We note that this is similar but not identical to the equation given by
Jackson et al. (2009).

The evolution of the semi-major axis is coupled to the evolution of the 
eccentricity, for which we use the equation of Jackson et al. (2009),
\begin{align}\label{jackson2}
\nonumber \frac{1}{e} \frac{de}{dt}= 
& - 
\left[ 
  \frac{63}{4} (G M^3_{\ast})^{1/2} \frac{R^5_{p}}{Q^{\prime}_{p} M_{p}}
\parenthnewln 
+ \frac{225}{16} \left(\frac{G}{M_{\ast}}\right)^{1/2} 
       \frac{R^5_{\ast} M_p}{Q^{\prime}_{\ast}}
  \right] a^{-13/2} 
\end{align} 

\begin{figure}
\epsscale{.80}
\plotone{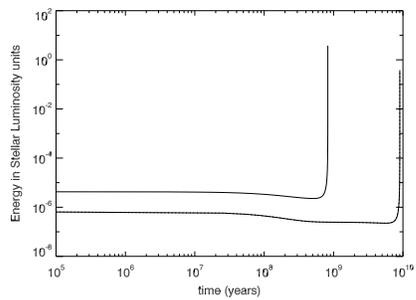} 
\caption{Energy output of the tides on the star during evolution of the semi-major for XO-3b for $Q_{\ast}=10^7$ and $Q_{\ast}=10^8$. Note the time will scales linearly with between values of $Q_{\ast}$. }\label{lumincr}
\end{figure}

To find the energy input into the star, 
we follow the derivation by Murray \& Dermott (1999) that this change
in energy as a function of the change in semi-major axis can be expressed as
the change in the orbital energy $dE_{orb}/dt$ minus the increase in
the rotation energy of the star $dE_{\ast,rot}$, and where we ignore the much
smaller increase in rotational energy of the planet as the orbital frequency
increases,
\begin{align}\label{dissipation1}
  \frac{dE_{orb}}{dt}-\frac{dE_{\ast,rot}}{dt} =
    \frac{d}{da}   \left[ \frac{-G(M_{\ast} M_p)}{2 a}
        -\frac{1}{2}I \omega_{\ast}^2 
  \right] \frac{da}{dt},
\end{align}
The result can be simplified by expressing it in terms of the mean motion,
\begin{align}\label{meanmotion}
n=\sqrt{\frac{G(M_{\ast}+M_p)}{a^3}} ,
\end{align} 
giving
\begin{align}\label{edissipation}
  \frac{dE}{dt}=- \frac{1}{2} \frac{M_{\ast} M_p}{(M_{\ast}+M_p)}
      n a (\omega - n) \frac{da}{dt},
\end{align} 
where in the case of an inwardly migrating planet spinning up a star 
we have $da/dt$ and $(\omega - n)$ less than zero. This loss in rotation energy
$dE/dt$ goes into the star. 
We show in Figure~\ref{lumincr} the amount of energy input into the star in units of stellar energy, using as an example the XO-3 system in which the mass of XO-3b is 11.8 $M_J$. We show that even for a stellar $Q_{\ast}$ value of $10^8$, during the last stages of tidal migration for the most massive planets the energy input into the star could rival the luminosity of the star. 
However, if the recent consensus (Lanza 2010; Barker \& Ogilvie 2010) 
that $Q_{\ast}$
for F stars such as XO-3 (a type F5V star) is $10^9$ or higher is correct, then the tidal input energy may rise to a less significant level. We will discuss the question ofthe whether rotational synchronization will end migration or if magnetic braking will slow the star to continue the planets' migration.
We leave for further work the important question of 
how fast this energy may be radiated away.

This paper focuses on actual observations of planet destruction due to tidal migration. We first briefly comment on previous work on evidence for planet destruction having already taken place, and on previous work on stars expanding and consuming their planets as a standard part of stellar evolution. This paper
empahsizes massive planets tidally infalling after planet formation
is complete, but we end with some comments on planet destruction while the circumstellar disk is still present.

\section{Tidal migration before and during destruction}
\subsection{Tidal migration rate coupled to stellar spin-up}
Migration into the star is coupled to the rotational spin-up of the star 
by the planet, so we need to follow and include the rotation of the star,
which is why we use the form for $da/dt$ (Equation~\ref{levardeq3})
of Neron de Surgy \& Laskar's (1997).
%

The slowdown of the rotation of the star as it sheds angular momenta 
from the system through ``magnetic braking'' is an important question for
further work. 
Equation (~\ref{levardeq3})  
shows that tidal migration is much more dependent on the angular motion of the planet relative to the star than on the planet's absolute rotational motion. The most massive planets have sufficient angular momentum to synchronize the rotation of the star, and in some cases 
(e.g. $\mathnormal{\tau}$ Boo) 
apparently may have done so. However, magnetic braking is expected to play such a significant role that even all synchronized star-planet systems may be expected to be unstable to merging, though how fast this will occur is of current theoretical work (Barker \& Ogilvie 2007).


Inward migration will be paused by RLOF, which causes outward migration (Chang et al. 2010; Gu et al. 2003; Paczynski 1966). Though estimates have been made (e.g. Gu et al. 2003), the equation of state of how much the inner planet will swell due to removal of  mass from the outside remains a less fully studied but important subject for planetary evolution. A delayed swelling of the inner planet  could lead to sustained RLOF even after the planet has started to migrate outward, leading to further outward migration. It will be important to look for signs alternating outward/inward patterns of migration (Chang et al. 2010) such as RLOF gas. It is important to determine how far the outward migration continues in order to determine whether currently known close-in planets may have already started down the planet destruction path. How strongly magnetic braking slows rotation is further complication by how the presence of the planet can disrupt the steady stellar wind, thus reducing the rate of angular momentum loss due to magnetic braking (Lanza 2010).

\subsection{Will migration for supermassive planets be stopped by synchronization?}

Levrard et al. (2009) raised the question of how many planets are unstable to
destruction versus how many planets will first synchronize the rotation of 
the star. Pont (2009) and others have begun the study of whether a correlation
between planet parameters and stellar rotation that might indicate such a
spin up. In part because of uncertainties in determining stellar ages 
the consensus is that the question be studied further.
Previous age determinations have used the stellar slow down in
rotation, but this cannot be used for planet hosting stars because
the rotation rate may have been altered by the planet.

\begin{figure}
\epsscale{.80}
\plotone{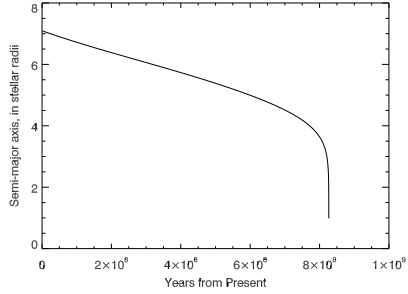} 
\caption{Evolution of the semi-major axis given in stellar radii for XO-3b for 
$Q_{\ast}=10^7$; the time will scale linearly with other values of $Q_{\ast}$. 
 The planet will inspiral much faster at 4 stellar radii and less, a distance 
at which the planet's angular momentum becomes too low to spin up the star 
enough to stop the inspiral.}\label{smaxisevol}
\end{figure}

To illustrate the evolution of a very high-mass planet, we use XO-3b as a good example system because its high mass of 11.8 $M_J$ (Winn et al. 2008) 
can be used to illustrate
the boundary between whether a massive planet is or is not stable to infall. 
Figure~\ref{starspinup} shows what would happen to the XO-3b system (with
period of 3.3 days  [Winn et al. 2008]) following 
Equation~\ref{levardeq3} if the star were to now be started with zero rotation,
where the frequency of the planet relative to the star is shown as a solid line
because it is what determines the migration.  This frequency is the 
difference between the orbital frequency (dashed line) and 
$\omega_{\ast} \cos \varepsilon $, the component the star's 
rotational frequency (dotted line) in the plane of the orbit. 
This illustrates that for such a massive planet that either the star's 
rotation would have to be very small or the orbit would have to be retrograde 
to not predict synchronization 
based on angular momenta alone. Rossiter-McLaughlin measurements have 
in fact found planets in retrograde orbits that would  satisfy this condition. 
The recent finding that many planets, especially the most massive planet 
systems likely including the XO-3 system (Winn et al. 2008), have retrograde
orbits means that there may be more massive planets that collide into
their stars rather become rotationally synchronized. It is clear from 
Figure~\ref{starspinup} that the contribution of the inclination is
likely the determining factor determining the eventual fate of these planets.
However, even if XO-3 were to have a prograde orbit, we can see that 
the region of stability afforded by synchronization could be ended if
magnetic braking were to slow the system's rotation to the point that 
the period goes under roughly 2 days. At this point it reaches a region 
in which the planet is so close to the star that its angular momentum is too
low to significantly spin up the star. This coincides with the stellar tides
becoming rapidly stronger, as shown by Figure~\ref{smaxisevol}, which shows 
the time evolution of the semimajor axis of the XO-3 system in which we
have used Equation 1 for zero initial stellar rotation. 
We see a final rapid inspiral due to the evolution equations being such
a stiff function of semimajor axis. 
Any planet that gets this close to a star will
rapidly spiral into the star, producing a luminous event.

\subsection{Stellar rotation slowed by magnetic braking.}

\begin{figure}
\epsscale{.80}
\plotone{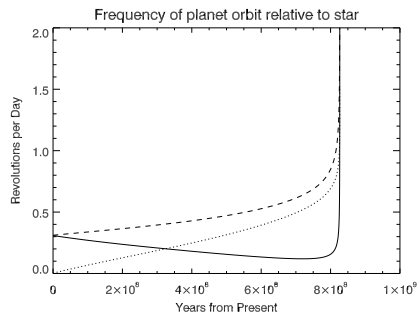} 
\caption{As the planet spirals in, angular momentum from the star is transferred from the planet to the star. We show what the evolution of the XO-3 system if the stellar rotation were initially at zero, by displaying the rotational frequency of the planet with a dashed line and the rotational frequency of the star as a dotted line. The all-important difference between the two, which can be thought of as the planet's rotational frequency in the star's reference frame, is shown as the solid line. If the star's initial rotation is high enough to make this line go to zero, as might be the case, the planet is then synchronized with the star.}\label{starspinup}
\end{figure}

Rotational synchronization becomes unstable if the planet's semi-major axis becomes too small, such as due to magnetic braking, as illustrated in the quick run-up of rotational frequencies in Figure~\ref{starspinup}. Figure~\ref{starspinup}, which shows the evolution of the semi-major axis as a function of time for the same parameters as Figure~\ref{starspinup}. We see that at too small of stellar radii, roughly below three stellar radii in this case, the planet simply has too little angular momenta to continue to spin up the star. A planet that has survived to this close distance is likely to see its orbit rapidly deteriorate through tidal migration, quickly ``falling'' into the star. Hence, it may be unlikely to find planets this close by usual planet-search methods due to the short amount of time planet may spend there, but the rapid energy release may provide a means of finding planets in a final doomed end state. Being so deep in the star's potential well (with consequent high orbital velocity) such a planet will release a very large fraction of the very large energies previously mentioned.

\subsection{RLOF and migration.}

The RLOF transfer of mass from the planet to the star causes outward migration of the planet due to conservation of angular momentum. Paczynski (1966) point out that the separation between two orbiting masses is smallest when their masses are equal. We follow Paczynski (1966) to show that the conservation of angular momentum requires that mass transfer from the planet to the star results in a larger semi-major axis.
We show the total angular momentum squared, $J^2$, of the planet-star system can be obtained relative to either body by the product of the velocity squared of the other body 
\begin{align}\label{vsquared}
v_2^2=\frac{G M_1^2}{a(M_1+M_2)}
\end{align} 
multiplied by the mass squared of the first body, $M_1^2$ multiplied by the separation, or semimajor axis, squared, $a^2$. Regardless of how we assign $M_{\ast}$ and $M_p$ to $M_1$ or $M_2$, we obtain, 

\begin{align}\label{totangmom}
J^2 = G a \frac{(M_{\ast} M_p)^2}{M_{\ast}+M_p} ,
\end{align} 
where $G$ is the constant of gravitation.  When angular momentum and mass
are conserved in the system, we have

\begin{align}\label{aChgWmass} 
a \left( M_{\ast} M_p \right)^2 =\frac{J^2 (M_{\ast}+M_p)}{G} = const.
\end{align}

We solve for a small change in semimajor axis $\delta a$ resulting
from a small transfer of positive mass gain to the star of 
$\delta M_{\ast}=\delta m$ from a mass loss from the planet of
$\delta M_p=-\delta m$. We obtain,

\begin{align}\label{migrationfrommassloss}
\delta a = \frac{2 a (M_{\ast}-M_p) \delta m}{M_{\ast} M_p},
\end{align}
which in the limit where $M_{\ast}+M_p \approx M_{\ast}$ becomes,
\begin{align}\label{migrationsmallmp}
\delta a \approx \frac{2 a \, \delta m}{M_p},
\end{align}
hence RLOF causes outward migration.

A major unanswered question is whether planets will continue to migrate inward as they RLOF until they are completely destroyed, or if the angular momentum transferred through RLOF will stall the inward migration and move them back outward, and if so, for how long? Will the RLOF merely pause the migration, or will a delayed rebound of the planet to the loss of mass move the planet outward? How many such inward/outward cycles occur before destruction? How much more rapid is the outward migration than the inward imgration? Might the planets known to be closest to their stars have already undergone an RLOF and outward migration event, or might we find one soon? In other words, what fraction of planets are undergoing this event? Gu et al. (2003) developed important theory that will be needed to address these questions.
A very massive planet may experience substantial changes in its equation of state through either a prolonged pause or prolonged cycles of alternating inward/outward migration (Chang et al. 2010). We suggest that a delayed RLOF could maintain an orbital eccentricity if more RLOF occurs after periastron than occurs before periastron, with the released gas giving the planet angular momentum transfer that is unbalanced through its orbit. Such a maintained eccentricity could enhance tidal heating of the planet, further increasing runaway RLOF. Perhaps runaway RLOF is sometimes a relatively fast event producing short-lived disks of expelled gas, if the runaway is not dampened by rapid outward migration. Runaway RLOF may be a very rapid event (Chang et al. 2010) leading to rapid energy release.

The structure of the released gas may be include a  disk (or ring) around the star, and/or a funnel from the planet towards the star depending on whether the  mass release is enough to overcome the stellar magnetic fields (Chang et al. 2010). Planet-produced disks of gas should be searched for structure, time dependent change, and asymmetry. An asymmetric disk (or ring) could be a sign of structure that could contribute to more massive planets having higher eccentricities. Rocky planets may respond differently, given that the RLOF distance decreases with increasing density. Similarly, when a massive gas planet's higher density core is exposed, the remaining planet may migrate further in.

A particularly unstable situation may occur for planets of several 
Jupiter masses, in the regime where an increase in mass actually 
reduces the planet's radius. Removal of outer material may lead to runaway RLOF
as each layer of material removed reduces the RLOF radius of the planet leading to even more RLOF, until the planet has either migrated much farther out or the planet's mass is reduced to where more mass loss reduces its radius. Gu et al. (2003) describe how adiabatic expansion continues to send material out mostly
through the inner Lagrange point L1. Gu et al. (2003) find that
RLOF through the outer Lagrange point L2 is not necessarily exluded, though 
Li et al. (2009) find in their model of WASP-12 sufficient loss 
of material throug L1 such that material is not lost through L2.
Such RLOF can cause rapid outward 
migration of the planet. We have shown that donservation of angular momentum 
dictates that loss of mass towards the star will lengthen the semi-major 
axis (Paczynski 1966). This outward migration brings the Roche distance 
further from the planet to the point that without sufficient 
explansion of the planet the RLOF and outward migration will cease, 
allowing the (slower) inward tidal migration to resume. 
(If a circumstellar disk has not yet fully dissipated, the planet could be 
migrated out into the disk, or the disk could be otherwise 
affected.) Such an effect could even contribute to the pileup 
in planet semi-major axes, if some of the Jupiter-mass planets 
in the pileup are actually remnants of planets of many Jupiter masses
that have already migrated in and out one or more times. 
The star may be significantly brightened by the gas released 
during such an event, possibly causing some of the gas to be 
expelled outward and carry angular momentum out of the system. 
Such an event may have observable signatures, either through 
luminosity variations or by signatures of a release of gas.

\subsection{Light curve signature of planet destruction}

Because planet destruction has only been modeled for limited cases 
\citep[e.g., ][]{sha88,sie99a,sie99a,sok03}, further work needs to be undertaken to identify the light curve that would be produced from stellar brightenings due to planet destruction. The results of these models shows that cases of the planet colliding into the star that the relevant time scales are on the order of one day, leading to our supposing what would be the observability of a planet destruction event if the orbital energy would be released in 24 hours. After considering the steps that planet destructon will go through, we conclude that the light curves may change dramatically with planet mass and other parameters. The light curve will follow from the steps of planet destruction, which may be summarized as follows:

\begin{itemize}
\item Input of tidal energy into star.
\item RLOF material falling into star.
\item Collision with photosphere will occur for more dense planets or cores when the Roche distance is less than the stellar photosphere.
\item Movement of material within star, which may vary from settling of material in the star, to continued orbit within the star (as in the case of expanded stars modeled by Siess \& Livio [1999a]).  This will release additional energy not included in calculations here.
\end{itemize}

Whether and how each step proceeds is highly mass dependent, and will also depend on the age and type of the star. Other factors including absorption by RLOF or whether RLOF affects the eccentricity need to be modeled.

\section{Infall of planetesimals during formation}

Planet formation models start with a higher number of planetesimals than planets in completed systems, and these planetesimals are presumed to frequently fall into the star (Trilling 1998; Chambers 2006). The energy available from planetesimal destruction by infall into the star is not nearly as great for much smaller objects, but the events are likely to be much more frequent, and recording such events could provide important constraints on planet formation models. Again, the important question is whether the energy release of such events might be diluted by slow RLOF, but perhaps the larger protostar might frictionally degrade the orbit to cause a more rapid infall to deeper into the protostar. RLOF is an important event, producing new disks, funnels, or perhaps strengthening the formation of jets of material from the protostar. The main question may be whether the presence of a disk might conceal either the energy or mass release either directly, or by how the fluctuations quantity of disk material changes both the environment near the star and the brightnesses of young stars. A large fraction of the photometric observations of accretion onto young stars by CoRoT (Alencar et al. 2010) show a regularity in their fluctuations that it is easy to imagine could be disrupted by a falling protoplanet.  In looking for exoplanetary system observables, it is important to learn whether planetesimal destruction produces observational signatures distinct from usual variations in the flow of disk material onto the star. Outburst behavior from the disk has long been studied (e.g., Zhu et al. 2010). Differentiating small protoplanet infall from such outbursts may be challenging since protoplanetary disk systems can have mass infall rates of $10^{-6}$ to $10^{-4} M_{\sun}$ per year.
Table 1 
shows how much energy could be released using typical solar system objects, for an assumed young sun radius of two solar radii.

\section{Future Work}
The full processes of planet destruction as a function of the properties of the planet and star remain to be modeled. It can be expected that the mass and composition of planets, and the mass and age of the star could all lead to different final outcomes for planets. It is important to model what happens to the migration of the orbit, how the planet's material is removed or if there is a collision, and what the effects are on the star. Though migration of the semi-major axis, along with consumption by evolving stars, will be the major routes to destroy planets, it can be asked if there will be planets made to collide with their stars by evolution in eccentricity or from planet-planet scattering.  The presence of a disk, and possibly other planets may affect this final and most active stage of a planet's evolution. Other planets and any disk may be affected by the energy output. 

Systems undergoing RLOF should be monitored for variation in luminosity, because RLOF processes may likely not be smooth. The gas will likely produce periodic negative and positive contributions to the system's luminosity. With planets possessing tens of thousands of years of stellar luminosity, a sufficiently rapid runaway mass loss could rival the star's luminosity. In the case of the WASP-12 system, Li et al. (2010) calculate that gas from the planet may intercept ~0.1 of the stellar visual luminosity, based on work on T Tauri disks (Adams et al. 1998; Hartmann et al. 1998). Based on this T Tauri disk work, Li et al. also report that viscous stress generates heat at a rate of $\sim 10^{-2} L_{\sun}$. We propose that observations be made not only to detect periodic negative or positive contributions to luminosity, but also to seek variations in what may be a turbulent process. This is yet one more reason to monitor transiting planet systems for variations in luminosity. We propose full-orbit continuous observations that at minimum could elucidate the structure of the released material and could provide atmosphere information on such misshapen planets, but may also reveal whether the release is a turbulent process. Repeated observations near transit, already worthwhile to seek transit timing variations, may also show variation in transit depth.

While this paper has been in preparation, much progress has been made in modeling (Gu et al. 2003; Trilling et al. 1998; Li et al. 2010; Lai et al. 2010) and observing (Fossati et al. 2010) the RLOF of WASP-12b. The models showed that the gas detected away from the planet is not merely being radiatively removed as is the case for HD 209458b, but is actually coming off through RLOF through an L1 Lagrange point nozzle. Chang et al. (2010) have modeled a giant planet undergoing RLOF in the presence of a protoplanetary disk. The obvious next step is to extend these models to planets similar to close-in planets being found by current transit surveys that do not have known protoplanetary disks. Understanding the response of planets' internal equation of state to the removal of outer layers is essential to knowing how angular momentum transfer changes the migration. Future statistics of the migration status of close-in planets will provide a handle on understanding planetary interior equation of states just as current work is placing limits on the values of planetary and stellar $Q_{\ast}$ values (Jackson et. al. 2010, in preparation; Barnes \& Taylor 2010, in preparation). It will also be important to couple successful stastical models of parameter distribution functions (e.g., Jiang et al. 2010; Jiang et al. 2009; 
 Jiang, et al. 2007) with events such as planet destruction that determine these patterns. It is to be expected that there will soon be many general models describing these events given the current high increase in the statistics of such close-in planets.

\section{Conclusions}
Searches for transiting planets are now finding close-in planets that have probably already begun the process of being destroyed. When we observe these planets we are seeing tidal migration in its most active state, with planets subjected to the highest force and subjected to the fastest sustained change. By observing planets in their final stage we will learn much about the properties of planets. The mass and energy releases accompanying planet destruction

We present the energies available to planets if their material is taken from RLOF distance to the surface of the star, and demonstrate the significance of this energy by comparing it to the amount of time for the star to radiate an equivalent amount of energy. We note that planet destruction energies compare with the energies of nova, and that nova are observed even from nearby galaxies. We promote how planet destruction should be considered a potential object of interest for upcoming transient surveys. Because these surveys already plan to do the ideal sort of observations to find large energy releases, it is important to characterize how this energy release may occur. 
We stress how that this energy release may occur at an unknown rate. RLOF may occur fast or slow, and there are cases where a planet may come in contact with the stellar photosphere, giving the fast energy release of a ``collision''. 

Planet destruction, however it occurs, offers a new phase space of opportunity for the study of planets. The manner of the energy release, as well as the release of planetary matter, will allow planets and stars to be observed in a manner that will illustrate properties that we could study in no other way.

\acknowledgments

Acknowledgements: We are grateful for the suggestions and support of Nader Haghighipour, Dan Fabrycky, Bruce Gary, Ed Guinan, Brian Jackson, Ing-Guey Jiang,
Richard Greenburg, Rory Barnes, Peter McCullough, Andy Gould, Scott Gaudi. The majority of this work was done under self support under the name ``Global Telescope Science,'' but we express gratitude because 
 proper completion could not have been 
appropriately achieved without the generous support of the National Tsing-Hua University, and the generous comments of the Institute of Astronomy faculty and staff. The Pan-STARRS Taiwan (PS-TW), the National Science Council of the Taiwan Grant NSC 99-2811-M-007-033 are gratefully acknowledged.

It is essential to claim that the majority part of this work was done under support of personal savings during extended unemployment, making a far-beyond 
normal need to acknowledge countless family and friends. Appreciation is given that colleagues helping with final comments encouraged claiming the earlier personal support. It is often said that without the support of the institution that this work would have been impossible. Indeed this work was much more difficult, but this prompts this unusual level of deep gratitude for all replacement support. Without the daily presence of colleagues, this work had repeatedly headed in wrong directions, so it is essentail to express a much higher than normal level of thanks for all those who remotely reviewed the many drafts. Support by the California Unemployment Fund for extended benefits was essential. This situation required such unusual support of the companion of the primary author that gratitude must be given to Emily Chang for unusual support and endurance.  
This paper is dedicated to the proposition
that astronomy institutes are obliged to provide participation 
that has been earned by any member making sacrifices in support of
the group effort. 
The support and cooperation of many unnamed parties to 
regain this participation is appreciated.
More cooperation is sought from coauthors. The personal contributions of many scientists is gratfully acknowledged. Appreciation is given to the European Science Foundation, the NASA Exoplanet Science Center, the American Astronomical Society, and the astronomers responding to an appeal for financial support for meeting attendance during a period of premature unemployment. The colleagues who
gave encouragement to claim credit for personal support are appreciated. 

The name ``Global Telescope Science (GTS)'' is given to the author's project to maintain participation in global telescope and exoplanet transit projects and research. 

Appreciation is given to astronomers at the following institutions for hosting in one form or another previous presentations of this work: University of Arizona, Arizona State University, New Mexico State University, University of Texas at Austin, Ohio State University, Johns Hopkins University, Harvard University, Pennsylvania State University, the University of Hawaii at Manoa, the Bishop Museum of Honolulu, and the University of Hawaii at Hilo.
Gratitude is also expressed to more amateur astronomy clubs that can be mentioned, but the help of members of the Santa Barbara Astronomical Unit stands out. 
Support by members of couchsurfing.org and strangers to allow the author to 
``sleep on couches'' in order to attend conferences is gratefull acknowledged.
This project has been improved by the practice in doing public presentation
affording by Toastmasters International, and the 
support of this club members' was 
essential. It was significantly helpful to have the support of organizers of
standup comedy events to help to see this work and 
related situations in the humorous light
that comes from presenting 
``an unemployed astronomer researching planet destruction'' in a novel venue. 
The encouragement of fellow comics that acknowledgement of comedy 
in a scientific paper could better promote both 
science and comedy is appreciated, as this could provide good lines 
of comedy useful to further promotion.
This paper was motivated to campaign to return to the areas of astronomy represented by the author's public talks of 2007-2010. 

Extensive use of the 
Extrasolar Planets Encyclopedia 
ADS, 
the Exoplanets Transit Database 
the Exoplanets Data Explorer 
the Amateur Data Archive 
the NASA Star and Exoplanet Database (NSTED),
and the astro-ph preprint server supported this research.

\clearpage 

\clearpage 

\clearpage 

\clearpage 

\clearpage 

\clearpage 

\clearpage 

\clearpage 







\clearpage

\begin{table}
\begin{center}
\caption{Energy release at RLOF distance relative to 2 solar radii for 
representative solar system bodies. The change in magnitude 
relative to the sun for
a one-day release of this energy is listed, followed by the absolute 
magnitude that such a 24-hour release of energy would give the sun.
\label{tbl-1}}
\begin{tabular}{crrrrrrrrrrr}
\tableline\tableline
Object & Mass & Radius     & Energy     & Energy   & Change in & Absolute  &\\
   &$M_{\sun}$& $R_{\sun}$ & Years of   & Days of  & Magnitude & Magnitude &\\
    &         &            & $L_{\sun}$ & $L_{\sun}$           &           &\\
\tableline
Mercury  & $1.7 \times 10^{-7}$ & $3.5 \times 10^{-3}$ & 1.9 & 480 & 6.7 & -1.9 \\ 
Venus    & $2.5 \times 10^{-6}$ & $8.7 \times 10^{-3}$ & 28 & 7000 & 9.6 & -4.9 \\ 
Earth    & $3.0 \times 10^{-6}$ & $9.2 \times 10^{-3}$ & 34 & 8600 & 9.8 & -5.1 \\ 
Mars     & $3.3 \times 10^{-7}$ & $4.9 \times 10^{-3}$ & 3.3 & 930 & 7.4 & -2.7 \\ 
Jupiter  & $9.6 \times 10^{-4}$ & $1.0 \times 10^{-1}$ & 6600 & $2.7 \times 10^{6}$ & 16.1 & -11.3 \\ 
Saturn   & $2.9 \times 10^{-4}$ & $8.7 \times 10^{-2}$ & 1600 & $8.2 \times 10^{5}$ & 14.8 & -10.0 \\ 
Uranus   & $4.4 \times 10^{-5}$ & $3.7 \times 10^{-2}$ & 300 & $1.2 \times 10^{5}$ & 12.7 & -8.0 \\ 
Neptune  & $5.2 \times 10^{-5}$ & $3.6 \times 10^{-2}$ & 390 & $1.5 \times 10^{5}$ & 12.9 & -8.2 \\ 
Pluto    & $6.6 \times 10^{-9}$ & $1.7 \times 10^{-3}$ & 0.1 & 19 & 3.2 & 1.5 \\ 
Moon     & $3.6 \times 10^{-8}$ & $2.5 \times 10^{-3}$ & 0.4 & 100 & 5.1 & -0.3 \\ 
Ganymede & $7.5 \times 10^{-8}$ & $3.8 \times 10^{-3}$ & 0.6 & 210 & 5.8 & -1.1 \\ 
Europa   & $2.4 \times 10^{-8}$ & $2.3 \times 10^{-3}$ & 0.2 & 69 & 4.6 & 0.1 \\ 
Titan    & $6.9 \times 10^{-8}$ & $3.7 \times 10^{-3}$ & 0.5 & 190 & 5.7 & -1.0 \\ 
Rhea     & $1.2 \times 10^{-9}$ & $1.1 \times 10^{-3}$ & $8 \times 10^{-3}$ & 3 & 1.6 & 3.2 \\ 
Enceladus& $6.0 \times 10^{-11}$& $3.7 \times 10^{-4}$ & $4 \times 10^{-4}$ & 0.16 & 0.2 & 4.6 \\ 
Ceres    & $4.8 \times 10^{-10}$& $7.3 \times 10^{-4}$ & $4 \times 10^{-3}$ & 1.4 & 0.9 & 3.8 \\ 
\tableline
\end{tabular}
\tablecomments{Table \ref{tbl-1} The amount of energy that would be 
released for gas falling from RLOF to a typical protestellar radius.}
\end{center}
\end{table}\label{protoplanettable}





\end{document}